\documentclass[letterpaper, 10 pt, conference]{ieeeconf}  

\IEEEoverridecommandlockouts                        

\usepackage{bm}
\usepackage{amsmath}
\usepackage{subfigure}
\usepackage{algorithm}
\usepackage{algorithmic}
\usepackage{booktabs}
\usepackage{color}
\usepackage{verbatim}
\usepackage{setspace}
\usepackage[breaklinks,colorlinks,linkcolor=black,citecolor=black,urlcolor=black]{hyperref}
\usepackage{multirow}

\usepackage{soul}

\usepackage[]{graphicx}    
\newcommand{\ignore}[1]{}  
\usepackage{bm}
\usepackage{mathrsfs}
\usepackage{amssymb}

\begin{document}

%
\title{Prediction-Based Reachability Analysis for  Collision Risk Assessment on Highways}
\author{Xinwei Wang$^{1}$, Zirui Li$^{1,2}$, Javier Alonso-Mora$^{3}$, Meng Wang$^{1,4}$
\thanks{This research was supported by the SAFE-UP project (proactive SAFEty systems and tools for a constantly UPgrading road environment). SAFE-UP has received funding from the European Union’s Horizon 2020 research and innovation programme under Grant Agreement 861570.}
\thanks{$^{1}$Xinwei Wang is with the Department of Transport and Planning, Delft University of Technology, Delft 2628 CD, The Netherlands
        {\tt\small x.w.wang@tudelft.nl}}%
        \thanks{$^{1,2}$Zirui Li is with the Department of Transport and Planning, Delft University of Technology, Delft 2628 CD, The Netherlands, and  also with the with the School of Mechanical Engineering, Beijing Institute of Technology, Beijing 100081, China.
        {\tt\small z.li@bit.edu.cn}}%
\thanks{$^{3}$Javier Alonso-Mora is with the Department of Cognitive Robotics, Delft University of Technology, Delft 2628 CD, The Netherlands
        {\tt\small j.alonsomora@tudelft.nl}}%
\thanks{$^{1,4}$Meng Wang is with      
        Chair of Traffic Process Automation, {Technische Universität Dresden}, {{Dresden}, {Germany}}, and with the Department of Transport and Planning, Delft University of Technology, Delft 2628 CD, The Netherlands,
        {\tt\small meng.wang@tu-dresden.de}}%
}


\maketitle

\begin{abstract}
\noindent Real-time safety systems are crucial components of intelligent vehicles. This paper introduces a prediction-based  collision risk  assessment approach on highways. Given a point mass vehicle dynamics system, a stochastic forward reachable set considering two-dimensional motion with vehicle state probability distributions is firstly established. We then develop an acceleration prediction model, which provides multi-modal probabilistic acceleration distributions to propagate  vehicle states. The collision probability is calculated by summing up the probabilities of the states where two vehicles spatially overlap. Simulation results show that the  prediction model has superior performance in terms of   vehicle motion position errors, and the proposed collision detection approach is agile and effective to identify the collision in cut-in crash events. 

\end{abstract}

\IEEEpeerreviewmaketitle

\section{Introduction}
Road traffic safety has attracted continuously increasing research attention, in particular in the current transition from conventional human-driven vehicles to automated and connected vehicles~\cite{mullakkal2020probabilistic}. To avoid potential vehicle crashes, extensive works on real-time collision detection have been conducted~\cite{mukhtar_vehicle_2015}. 

Collision detection can be generally divided into three methodologies, i.e., neural network-based approaches, probabilistic approaches, and formal verification approaches. 
Neural networks have potential to provide accurate vehicle collision detection through classifying safety-critical scenarios. For instance, a collision detection model using a neural network based classifier was developed in~\cite{katare_embedded_2019}. The proposed model takes on-board sensor data, including acceleration, velocity, and separation distance, as input to a neural network based classifier, and outputs whether alerts are activated for a possible collision. A specific kangaroo  detection approach was proposed in~\cite{saleh_kangaroo_2016}, where a deep semantic segmentation convolutional neural network is trained to recognize and detect kangaroos in dynamic environments. Although neural network-based approaches are effective to identify potential collisions, the trained classifier generally cannot include clear decision rules and is hard to interpret. 

To address uncertainties of surrounding vehicles, probabilistic based approaches have also been widely adopted for collision detection. A conceptual framework to analyze and interpret the dynamic traffic scenes was designed in~\cite{laugier_probabilistic_2011} for collision estimation. The collision risks are estimated as stochastic variables and predicted relying on driver behavior evaluation with hidden Markov models. A probability field for future vehicle positions was defined in~\cite{annell_probabilistic_2016}, where  an intention estimation and a long-term trajectory prediction module are combined to calculate the collision probability. 
Given a set of local path candidates, a collision risk assessment considering lane-based probabilistic motion prediction of surrounding vehicles was proposed in~\cite{kim_collision_2018}. However, these methods typically require pre-defined parameters of position distributions, which can  impact the adaptability of the probabilistic collision detection.

Another mainstream to address the collision detection is formal verification approaches~\cite{pek2017verifying,tornblom2019abstraction}, among which reachability analysis (RA) can compute a complete set of states that an agent (e.g. a vehicle) can reach given an initial condition within a certain time interval~\cite{althoff2021set}. Based on RA, a safety verification thus is performed by propagating all possible reachable space of the automated vehicle (AV) and other traffic participants forward in time and checking the overlaps. One major advantage of RA is that safety can be theoretically guaranteed if such forward reachable set (FRS) of the automated vehicle does not intersect that of other traffic participants for all times. 

The standard RA approach suffers from over-conservatism. To reduce the over-conservative nature of  forward reachability, a stochastic FRS discretizing the reachable space into grids with probability distributions was developed in~\cite{althoff2009model}. At each time step, a collision probability is provided by summing probabilities of the states that vehicles 
intersect. Then a collision threshold can be set to check whether the current vehicle interactions are safe or not. However, this approach is based on Markov chains, which assume that the vehicle state and its control input evolves only in line with  the current state. Besides, it cannot explicitly address two-dimensional motion, as lane-change maneuvers are not considered.

In this work, we propose a prediction-based collision detection approach on highways based on stochastic FRS, where the state probability distribution of each surrounding vehicle is obtained by leveraging a neural network-based acceleration prediction model. 
The main contribution is  the establishment of a stochastic FRS for each surrounding  vehicle considering two-dimensional positions and velocities to address two-dimensional motion uncertainties. The state transition probabilities are provided by a  long-short term memory (LSTM) model for acceleration prediction. The proposed acceleration prediction model has a two-stage structure, and its input features are selected and processed differently at each stage. The model is trained to minimize propagated vehicle position errors.

  

The remainder of the paper is organized as follows: Section~\ref{sec:pre} provide preliminaries on Markov-based stochastic FRS and the employed vehicle dynamics, and in Section~\ref{sec:models} we propose a prediction-based stochastic FRS on highways for collision detection. Simulations are conducted in Section~\ref{sec:sim} to verify the performance of the proposed collision detection approach. Finally, conclusions are drawn in Section~\ref{sec:conc}.

\section{Preliminaries}\label{sec:pre}

\subsection{Markov-based stochastic FRS}

In this work, we use notations from~\cite{althoff2009model} with minor modifications to describe the Markov-based stochastic FRS. The computation of the FRS is done by considering all possible control inputs of a system given an initial set of states. 
The FRS of a system is formally defined as

\begin{equation}\begin{split}
    \mathcal{FR}(t,\bar{t}) := &\{ \chi   (\bar{t},x(t), u(\cdot)) | x(t)\in \mathcal{X}_{0}, \\ &\forall t^{\star} \in [t,t+\bar{t}] : \chi (t^{\star},x(t), u(\cdot))) \in \mathcal{X}, u(t^{\star}) \in \mathcal{U}
    \}
    \end{split}
\end{equation}
where $x$ is the system state, and $\mathcal{FR}(t,\bar{t})$ is a forward set that the system is reachable at time $t+\bar{t}$ from an initial set $\mathcal{X}_{0} \subset \mathcal{X} $ at time $t$ and subject to any input~$u$ belonging to the admissible control input set~$\mathcal{U}$.

One of the most frequently used techniques is to approximate stochastic processes by Markov chains, 
 which present a stochastic dynamic system with discrete states
 ~\cite{althoff2009model}.  The discretized future time series are denoted as $t+t_{k} (k\in \{1,\dots,e\})$, where $e$ is the future final time step, and the duration of the time step is $dt$. Due to the stochastic characteristics, the system state at predicted time step is not exactly known, and a probability $p_{i}(t+t_{k})$ is assigned to each state $i$ at the current time $t+t_{k}$. Then the probability vector $\mathbf{p}(t+t_{k+1})$ composed of probabilities $p_{i}(t+t_{k})$ over all states is updated as
 

\begin{equation}
   \mathbf{p}(t+t_{k+1})=\Phi \cdot \mathbf{p}(t+t_{k}) \label{pTrans}
\end{equation}
where $\Phi$ is the state transition matrix. Here $\Phi$ is time invariant as the model is assumed as Markovian.

To implement a Markov chain model, the system state first needs to be discretized if the original system is continuous. For the vehicle dynamic system, we represent it as a tuple with four discretized elements, including two-dimensional vehicle positions and velocities. Meanwhile, the control input requires to be discretized. Detailed discretization parameters are reported in Section~\ref{sec:setup}.


Each element $\Phi_{ji}$ in matrix $\Phi$ represents the state transition probability from state $i$ to $j$. 
Note that the transition probabilities depend on the discrete input $u$ as well, i.e., each discrete input $u$ generates a conditional transition probability matrix $\Phi^{u}$. Specifically, each element $\Phi^{u}_{ji}$ in the conditional matrix $\Phi^{u}$ is the possibility starting from the initial state $i$ to $j$ under  acceleration $u \in \mathcal{U}$, where  $u$ represents the corresponding acceleration of $\Phi_{ji}^{u}$. The conditional probability $\Phi_{ji}^{u}$ therefore is expressed as 


\begin{equation} \label{eq:phi_u}
\Phi_{ji}^{u} =
\left\{
\begin{aligned}
        p_{i}^{u}& ,~\text{if state} \ {i} \ \text{reaches state} \ j \ \text{with input}\ u\\
        0& ,~\text{otherwise}
        \end{aligned}
        \right.
\end{equation}
where $p^{u}_{i}$ is the control input probability given state $i$. The time index does not appear here as it is a Markov process. The overall state transition matrix is then constructed as

\begin{equation}  \label{eq:phi}
   \Phi_{ji} = \sum\limits_{u \in\mathcal{U} } \Phi_{ji}^{u}
\end{equation}

The probability distribution of the control input $p^{u}_{i}$ is dynamically changed by another Markov chain with transition matrix $\Gamma_i$, depending on the system state $i$. This allows a more accurate modeling of driver behavior by considering the frequency and intensity of the changes of control input. As a consequence, the transition matrices $\Gamma$ have to be learned by observation or set by a combination of simulations and heuristics. By incorporating the two transition matrices $\Phi$ and $\Gamma$, a Markov-based stochastic FRS with probabilities $\mathbf{p}(t+t_{k})$ over all discretized states can be obtained at each predicted time step $k$.  

In~\cite{althoff2009model}, the acceleration (i.e., control input) transition probability matrices $\Gamma$ only depend on the acceleration and the state at the current time.  The computational efficiency is ensured by using such simplified Markovian setting, while the future acceleration and trajectories of a vehicle can be influenced by historical information~\cite{deo2018convolutional}. Therefore in this work, we aim to utilise a vehicle acceleration predictor with multi-maneuvering modes to generate and dynamically update the transition matrices.

\subsection{Vehicle Dynamics}

To compute the stochastic FRS of the surrounding vehicle, we adopt a point mass model and its control input  is expressed as two-dimensional accelerations with probabilistic bivariate normal distributions, which are predicted by a learning-based model that will be introduced in Section~\ref{sec:4.1}. {Here we use a simple point mass model, since the main position errors depend on the performance of the future control input prediction. The two-dimensional accelerations are also compatible with the existing control input prediction models~\cite{althoff2009model,fridovich2020confidence}}. Based on the point mass vehicle model, the future vehicle system states, which are discretized as a tuple of two-dimensional positions and velocities, can be propagated with the predicted accelerations at each time step.  The vehicle dimension size is to be considered when checking whether two vehicles collide. We assume that the planned trajectories of the ego vehicle are known in advance. {The uncertainties of  ego vehicle motions and road environments are not considered in this work,  while these can be modelled by extending its planned trajectories with a bound set. The ego vehicle could then occupy more states at each time step, leading to a higher collision probability due to uncertainties~\cite{mitchell2007toolbox}} .

 \section{Prediction-based stochastic FRS} \label{sec:models}
 
 In this section, to provide more accurate prediction of surrounding vehicles, we first introduce a two-stage multi-modal acceleration prediction model consisting of a lane change maneuver prediction module and an acceleration prediction module. Then we detail how the stochastic FRS is established through incorporating the proposed acceleration prediction model. 
 
 \subsection{Acceleration prediction of a surrounding vehicle}\label{sec:4.1}
 
   \begin{figure*}[!tb]
\label{sec:model}
	\begin{center}
		\includegraphics[width=0.7\textwidth]{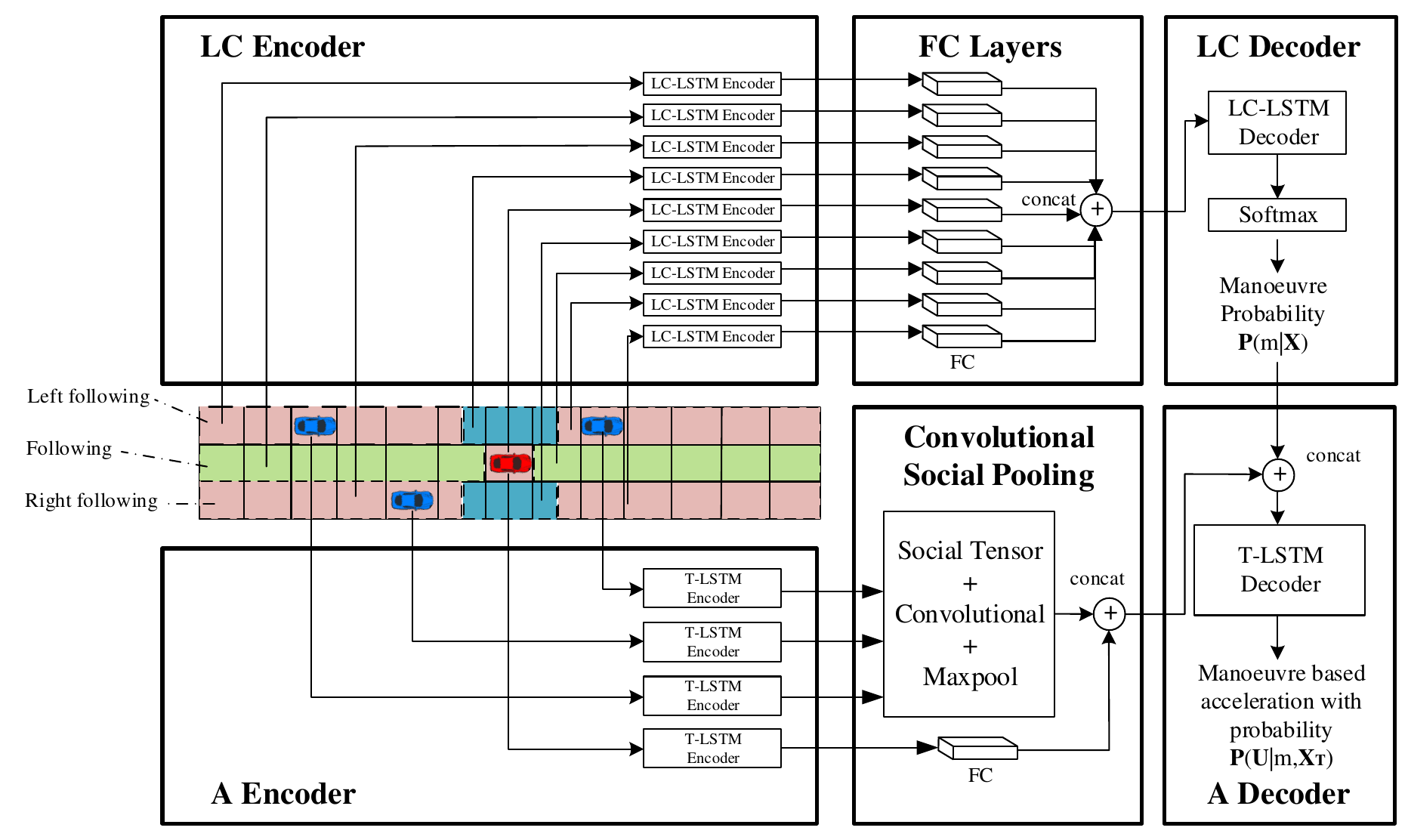}
		\caption{Overview of the acceleration prediction model,  consisting of a lane-change maneuver prediction module and an acceleration  prediction module (denoted as LC and A respectively in the figure). The two modules both have a encoder-decoder structure, but adopting and processing historical information as input in different ways. Abbreviations concat and FC stand for the concatenation operation and fully connect layer respectively. A
variant of the model for trajectory prediction was developed earlier in~\cite{wang2021probabilistic}.  }\label{fig:model}
		\end{center}
		\end{figure*}
		
 Existing works either use heuristic rules~\cite{althoff2009model} or action-state values~\cite{fridovich2020confidence} to represent future acceleration distributions of a surrounding object. However, these methods predict the object state and control input only with the current state. 
 Typically, the vehicle trajectories and acceleration are predicted using both current and historical information~\cite{deo2018convolutional}. In doing so, the prediction accuracy can be improved compared to that only using  current states as input. This motivates us to establish an LSTM based network to dynamically predict probabilistic future vehicle accelerations both using current and historic vehicle information. An overview of the developed two-stage acceleration prediction model is illustrated in Fig.~\ref{fig:model}.
 
 
 \subsubsection{Two-stage vehicle acceleration prediction}
We have developed a two-stage multi-modal trajectory prediction model in~\cite{wang2021probabilistic}. In this work,  we keep the same lane-change maneuver prediction model at the first stage, but develop a new acceleration prediction model at  the second-stage model. This is because that the acceleration prediction is employed to enable the dynamic update of the conditional probability $\Phi_{ji}^{u}(t+t_{k})$.

We first briefly introduce the adopted lane-change maneuver prediction module from~\cite{wang2021probabilistic}. The input of the module is expressed as 

\begin{equation}
\mathbf{X} = [\mathbf{x}^{(t-t_h)},\dots,\mathbf{x}^{(t-1)},\mathbf{x}^{(t)}]
\end{equation}
where $\mathbf{X}$ represents all input features from time $t-t_{h}$ to $t$. At each historic time step, the collected input is composed of three parts: $\mathbf{x}^{(t)} = [\mathbf{x_{T}}^{(t)},\mathbf{b}^{(t)},d^{(t)}]$, where $\mathbf{x_T}^{(t)}$ is the trajectory information for vehicle being predicted as well as its surrounding vehicles, $\mathbf{b}^{(t)}$ contains two binary values to check whether the predicted vehicle can turn left and right, and $d^{(t)} \in [-1,1]$ is the normalized deviation value from the current lane center.

As shown on the top of Fig.~\ref{fig:model}, LSTMs are used to encode and decode the lane-change maneuver prediction model, in which the encoding information is passed to fully connected layers before decoding. The output of the model is a probability distribution $\mathbf{P}(m|\mathbf{X})$ for each lane-change maneuver mode from time $t+1$ to $t+t_f$.

As for the acceleration prediction at the second stage, the input includes historic positions of the vehicle being predicted and surrounding vehicles, in addition to the historic accelerations $\mathbf{x_A}^{(t-t_h:t)}$ of the vehicle being predicted:
\begin{equation}
\mathbf{X_{T}} = [\mathbf{x_T}^{(t-t_h)},\dots,\mathbf{x_T}^{(t-1)},\mathbf{x_T}^{(t)}, \mathbf{x_A}^{(t-t_h:t)}]
\end{equation}

As we use additional acceleration information for the vehicle being predicted, we modify the input size of the LSTM encoder in~\cite{wang2021probabilistic} for the vehicle being predicted, while maintaining the overall network structure unchanged. Detailed information of the second-stage model is referred to~\cite{deo2018convolutional,wang2021probabilistic}.

Given the input $\mathbf{X_T}$ and corresponding maneuver mode probability distribution $\mathbf{P}(m|\mathbf{X})$, the output $\mathbf{P}(\mathbf{U}|m,\mathbf{X_T})$ of the second-stage acceleration prediction model is conditional acceleration distributions over 
\begin{equation}
\mathbf{U} = [\mathbf{u}^{(t+t_{1})},\dots,\mathbf{u}^{(t+t_e)}]
\end{equation}
where $\mathbf{u}^{(\cdot)}$ is the predicted vehicle acceleration at each time step within the prediction horizon. 
Note that the prediction horizon and time increment are the same as those for the reachable set computation, respectively.

Given the three defined maneuvers $m$, the probabilistic multi-modal distributions are calculated as 
\begin{equation}
\mathbf{P}(\mathbf{U}|\mathbf{X,X_{T}})  = \sum_{m} \mathbf{P}_\Theta (\mathbf{U}|m,\mathbf{X_T})\mathbf{P}(m|\mathbf{X})
\end{equation}
where outputs $\Theta = [\Theta^{(t+t_{1})},\dots,\Theta^{(t+t_e)}]$ are time-series bivariate normal distributions. Specifically $\Theta^{(t+t_{k})} = \{\mu_{1m}^{k},\mu_{2m}^{k},\sigma_{1m}^{k},\sigma_{2m}^{k},\rho_{m}^{k}\}_{m=\{1,2,3\}}$ corresponds to the predicted acceleration means and standard deviations along two dimensions, and the correlation at  future  time  instant $t+ t_k$ under each maneuver mode $m$, respectively. 

Under acceleration distributions $\Theta$, the future vehicle trajectories are propagated as 
\begin{equation}\label{eq:propagate}
\left \{
\begin{aligned}{}
    v_{1m}^{k+1} =  &  v_{1m}^{k} + \mu_{1m}^{k}dt \\
    v_{2m}^{k+1} =  &  v_{2m}^{k} + \mu_{2m}^{k}dt\\
    y_{1m}^{k+1} =  &   y_{1m}^{k}  + (v_{1m}^{k+1} + v_{1m}^{k})dt/2 \\
    y_{2m}^{k+1} =  &   y_{2m}^{k}  + (v_{2m}^{k+1} + v_{2m}^{k})dt/2
\end{aligned}
\right.
\end{equation} 
where $dt$ is the time increment, $v_{1m}^{k}, v_{2m}^{k}, y_{1m}^{k},y_{2m}^{k}$ are the propagated two-dimensional velocities and positions at  future  time  instant $t+ t_k$ for each maneuver mode $m$, respectively. $(v_{1m}^{0}, v_{2m}^{0}, y_{1m}^{0},y_{2m}^{0})$ denotes the system state at the current time $t$. 
The propagated trajectory variances are updated as $\widetilde{\sigma}_{1m}^{k} = {\sigma}_{1m}^{k}\cdot(dt)^{2}/2$ and $\widetilde{\sigma}_{2m}^{k} = {\sigma}_{2m}^{k}\cdot(dt)^{2}/2$, and the correlation remains the same as $\rho_{m}^{k}$. Therefore, the propagated probabilistic distributions  of the vehicle position are expressed as $\widetilde{\Theta}^{(t+t_{k})} =\{y_{1m}^{k},y_{2m}^{k},\widetilde\sigma_{1m}^{k},\widetilde\sigma_{2m}^{k},\rho_{m}^{k}\}_{m=\{1,2,3\}}$\@.

 \subsubsection{Model training}
 Typically a multi-modal prediction model is trained to minimize the negative log likelihood (NLL) of its conditional distributions as
 
\begin{equation} \label{eq:nll}
-\text{log}\left(\sum_{m} \mathbf{P}_\Theta (\mathbf{U}|m,\mathbf{X_{T}})\mathbf{P}(m|\mathbf{X})\right)
\end{equation}
For more accurate collision probability estimation,  we focus on the potential  collision when two vehicles have intersections along the trajectories. We therefore directly minimize the trajectory prediction errors propagated from the acceleration prediction in line with~\cite{zhou2017recurrent} as 
\begin{equation} \label{eq:nll-revised}
-\text{log}\left(\sum_{m} \mathbf{P}_{\widetilde{\Theta}} (\mathbf{Y}|m,\mathbf{X_{T}})\mathbf{P}(m|\mathbf{X})\right)
\end{equation}
where $\mathbf{Y}= [\mathbf{y}^{(t+1)},\dots,\mathbf{y}^{(t+t_f)}]$ is the propagated trajectories with distributions $\widetilde{\Theta}$, and $\mathbf{y}^{(k)} = \{y_{1m}^{k},y_{2m}^{k}\}$ are the predicted positions of the vehicle at time step $k$ under maneuver mode $m$.

To further improve the prediction performance, we separately train the lane-change maneuver and vehicle acceleration prediction models. This is because that the proposed approach has a two-stage structure: the maneuver probabilities are first predicted, and then for the corresponding conditional vehicle acceleration distributions.   For the maneuver prediction model, it is trained to minimize the NLL  of the  maneuver probabilities $-\text{log}\left(\sum_{m} \mathbf{P}(m|\mathbf{X})\right)$; for the vehicle acceleration prediction, the adopted model is to minimize $-\text{log}\left( \sum_{m} \mathbf{P}_{\widetilde{\Theta}} (\mathbf{Y}|m,\mathbf{X_T})  \right)$.
 
 \subsection{Prediction-based stochastic FRS of a surrounding vehicle}
When predicting future states of a surrounding vehicle, not only the current state but also historical information needs to be considered~\cite{deo2018convolutional}. In this work, 
we use the acceleration prediction results from Section~\ref{sec:4.1} to dynamically update the state transition probability matrix at each time step.

 
 The system state $i$ of the surrounding vehicle is represented as a tuple with four discretized elements, including two-dimensional vehicle positions and velocities. The system input is expressed as a two-dimensional acceleration $(a_{1},a_{2})$. Note the current state probability is known in advance. Typically there is an initial state $i$ with $p_{i}(t_{0})=1$, or an initial probability distribution is provided to address state uncertainties. In practice, from the current time $t$, we need to calculate multiple stochastic FRSs at multiple forwarded time steps, and check the corresponding FRS at each future time step $k \in \{ 1,2,\dots,e\}$. 
 
 At each predicted time step $k$,  the acceleration prediction model provides a bivariate normal distribution  function $f_{m}^{k}(a_{1},a_{2})$ for each maneuver mode $m$ as

\begin{equation}
\scriptsize
\begin{split}
&f^{k}_{m} (a_{1},a_{2})= \frac{1}{2\pi\sigma_{1}\sigma_{2}\sqrt{1-\rho^{2}}}\cdot \\ &\text{exp} \left( -\frac{1}{2(1-\rho^{2})} \left[\left ( \frac{a_{1}-\mu_{1}}{\sigma_{1}}\right)^2 + \left ( \frac{a_{2}-\mu_{2}}{\sigma_2}\right)^2 - 2\rho\frac{(a_{1}-\mu_{1})(a_{2}-\mu_{2})}{\sigma_{1}\sigma_{2}}\right ]  \right) 
\label{eq:PDF}
\end{split}
\end{equation}
where $\mu_{1},\mu_{2},\sigma_{1},\sigma_{2},\rho$ provided by the prediction model  denote predicted means and standard deviations along two directions, and the correlation at  future  time  instant $t+ t_k$ for each maneuver mode $m$, respectively. The time and maneuver indices of the five parameters are omitted here for the sake of brevity.

 To propagate the system states, the conditional probability $p^{u}_{i}(t+t_{k})$ at time step $k$ under state $i$ and acceleration $u = (a_{1}^{u},a_{2}^{u})$ is calculated as 
 
 \begin{equation} \label{eq:u}
  \scriptsize
    {p}_{i}^{u}(t+t_k) = \frac{\widetilde{p}_{i}^{u}(t+t_k)}{\sum\limits_{u \in\mathcal{U}  } \widetilde{p}_{i}^{u}(t+t_k) }
 \end{equation}
 
 \begin{equation} \label{eq:u-assist}
 \scriptsize
     \widetilde{p}_{i}^{u}(t+t_k) =  \sum\limits_{m} \lambda_{m}^{k}\cdot \int_{\underline{a}_{2}^{u}}^{\overline{a}_{2}^{u}} \int_{\underline{a}_{1}^{u}}^{\overline{a}_{1}^{u}} f_{m}^{k}(a_{1},a_{2}) da_{1} da_{2}
 \end{equation}
 where $\lambda_{k}^{m}$ is the probability for maneuver mode $m$ at time step $k$, and $\underline{a}_{1}^{u},\overline{a}_{1}^{u},\underline{a}_{2}^{u},\overline{a}_{2}^{u}$ are the integral boundaries of $u$.
 
 Here the conditional state probability $p^{u}_{i}(t+t_k)$ is implicitly relevant to the current state as well as historical states. This is because the current and historical information has been considered when providing the predicted acceleration results. This implies the state transition  matrix now has to be computed online. 
 
 Substituting~\eqref{eq:u} and~\eqref{eq:phi_u} into~\eqref{eq:phi}, the overall state transition matrix $\Phi$ is obtained. To distinguish the Markov-based approach which can compute the transition matrix offline, we denote the state transition matrix obtained with the prediction model at the predicted time step $k$ as $\Phi(t+t_{k})$. Then at each predicted time step, the state probability vector is iteratively computed as 
 
  \begin{equation} \label{eq:transition-t}
    \mathbf{p}(t+t_{k+1}) = \Phi(t+t_{k})\cdot \mathbf{p}(t+t_{k})
 \end{equation}

To measure the driving risk, the collision probability at the current time $t$ is expressed as the product of collision  probability at each predicted time step:
  \begin{equation}
  \scriptsize
        P_{col}(t) = 1 -\prod \limits_{k} \left (1-\sum\limits_{i \in  \mathcal{H}(t+t_{k})}p_{i}(t+t_{k})\right)
 \end{equation}
 where $\mathcal{H}(t+t_{k})$ is the set of states that  the ego vehicle position occupies at time step $k$. The vehicle dimension is considered when calculating the collision probability. 


\section{Simulations}\label{sec:sim}

\subsection{Dataset and setup}\label{sec:setup}

The highD dataset~\cite{krajewski2018highd}, which contains bird-view naturalistic driving data on German highways, is utilized to train and test the acceleration prediction model. We randomly select equal samples for the three different lane-change maneuver modes, leading to 
135,531 (45,177 for each maneuver mode) and 19,482 (6,494 for each mode) samples for the training and testing respectively. The original dataset sampling rate is 25 Hz, and we downsample by a factor of 5 to reduce the model complexity. We consider 2-seconds historic information as input and predict within a 2-second horizon.  

The prediction model is trained using Adam with learning rate 0.001, and the sizes of the encoder and decoder are 64 and 128 respectively. The size of the fully connected layer is 32. The convolutional social pooling layers consist of a $3\times 3$ convolutional layer with 64 filters, a $3\times 1$ convolutional layer with 16 filters, and a $2\times 1$ max pooling layer, which are consistent with the settings in~\cite{deo2018convolutional}.

The vehicle longitudinal (lateral) positions are discretized from -2 to 80 (-4 to 4) meters with an increment 2 (1) meters,  and the longitudinal (lateral) velocities are discretized from 20 to 40 (-2.5 to 2.5) m/s with an increment 0.4 (0.2) m/s, leading to around half a million states. As for the control input, we discretize the longitudinal (lateral) accelerations  from -5 to 3 (-1.5 to 3) m/s$^2$ with an increment 1 (0.5) m/s$^2$, leading to 63 acceleration combinations. We also add several constraints to limit the acceleration selection, including maximal acceleration, strict forward motion, and maximal steering angle~\cite{mullakkal2020probabilistic}. In the end, 37 million possible state transfers are generated. To alleviate the computational load, we assume that an advanced GPU~\cite{turner2018application}, which enables 2048$\times$28 parallel 
computation, is available. The stochastic FRS with state probability distributions $\mathbf{p}({t+t_{k}})$ is calculated at each predicted future time step within 2 seconds with an increment 0.4 seconds, i.e., $t_{k} \in \{0.4, 0.8, 1.2, 1.6, 2.0\}$.

\subsection{Results and discussions}

We first report the performance of the proposed prediction model (denoted as A-LSTM) and a baseline model SC-LSTM~\cite{deo2018convolutional}. SC-LSTM is an LSTM network with social convolutional layers, which has competitive performance for trajectory prediction.  Note that although A-LSTM is developed for acceleration prediction, we train the model to minimize the propagated vehicle position errors as~\eqref{eq:nll-revised} to provide accurate collision probability estimation. Therefore, we compare the two approaches in Table~\ref{tab:prediction_model} with respect to four evaluation indicators. RMSE, ADE, FDE  are the average root mean square error, displacement error, final displacement error of the future predicted motion positions respectively, and NLL is the negative log likelihood  the  of~\eqref{eq:nll}. A lower value of NLL corresponds to more accurate multi-modal prediction performance. Column Dif denotes the relative difference between SC-LSTM and A-LSTM.  We do not calculate the difference of NLL, as it does not make sense.

\begin{table}[htbp]
  \centering
  \scriptsize
  \caption{Prediction results between A-LSTM and SC-LSTM on the testing dataset.}
    \begin{tabular}{lrrr}
    \toprule
          & \multicolumn{1}{c}{SC-LSTM} & \multicolumn{1}{c}{A-LSTM} & \multicolumn{1}{c}{Dif(\%)} \\
    \midrule
    RMSE (m) & 0.596 & 0.215 & 63.93  \\
    ADE (m) & 0.333 & 0.125 & 62.46  \\
    FDE (m) & 0.859 & 0.335 & 61.00  \\
    NLL   & -0.822 & -3.071 & / \\
    \bottomrule
    \end{tabular}%
  \label{tab:prediction_model}%
\end{table}%

The proposed A-LSTM clearly has superior performance compared with the baseline approach SC-LSTM in terms of all evaluation indicators. This is mainly due to the two-stage network structure of A-LSTM, while CS-LSTM uses one network to simultaneously predict the lane-change maneuver mode and  the future vehicle positions.

When analyzing the trajectories in highD, it is found that almost all trajectories are not safety-critical, leading to zero collision probability, no matter which collision detection approach is employed. Consequently, it is hard to distinguish different collision detection approaches using scenarios/trajectories in highD. { The proposed collision detection approach is generally applicable to all scenarios on highways, while its advantages can be better exploited in safety-critical events.  Therefore, we simulate safety-critical cut-in trajectory data to test different collision detection approaches, since cut-in events are potentially risky on highways~\cite{liu2015classification}}. 

In the simulated cut-in event, the ego vehicle travels on the middle lane and the surrounding vehicle travels on the right lane with a constant longitudinal velocity 31 and 28 m/s, respectively. The surrounding vehicle is 15 meters ahead of the ego at $t = 1$ second, and starts to turn right with a constant lateral acceleration before crossing the lane marker at $t=4.8$ seconds. The car length and width of both vehicles are set as 4 and 2 meters, respectively. As the longitudinal velocity of the ego is greater than that of the surrounding one, a crash occurs at around $t=5$ seconds.

\begin{figure}[htb]
	\begin{center}
		\includegraphics[width=0.5\textwidth]{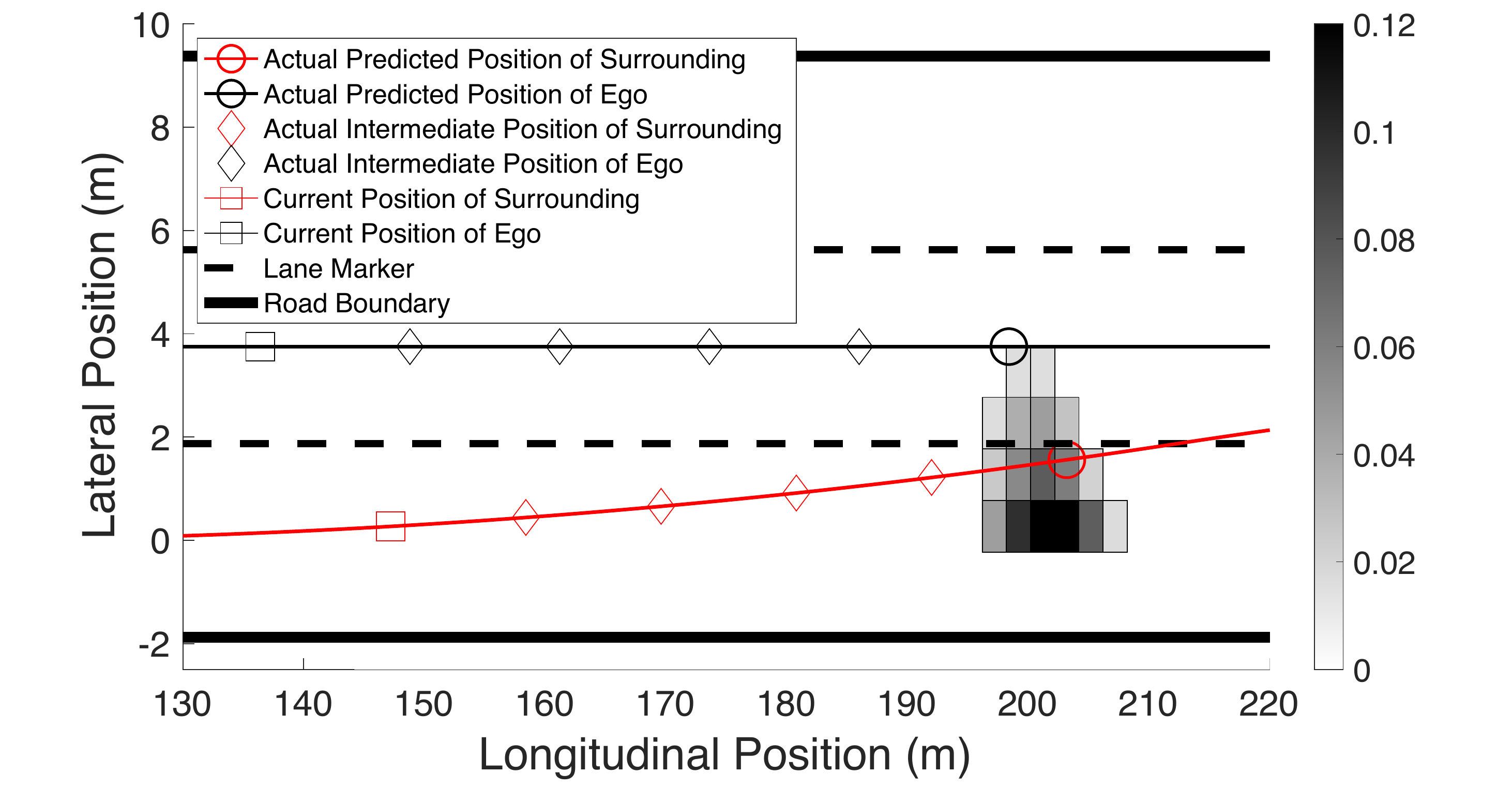}
		\caption{Visualized probability distributions  at the current time 2.4 seconds for the predicted time 4.4 seconds using the approach in~\cite{althoff2009model}. Only the positions with a probability greater than 1\% are displayed. Dark color indicates high distribution probability.}\label{fig:heu}
	\end{center}
\end{figure}

\begin{figure}[htb]
	\begin{center}
		\includegraphics[width=0.5\textwidth]{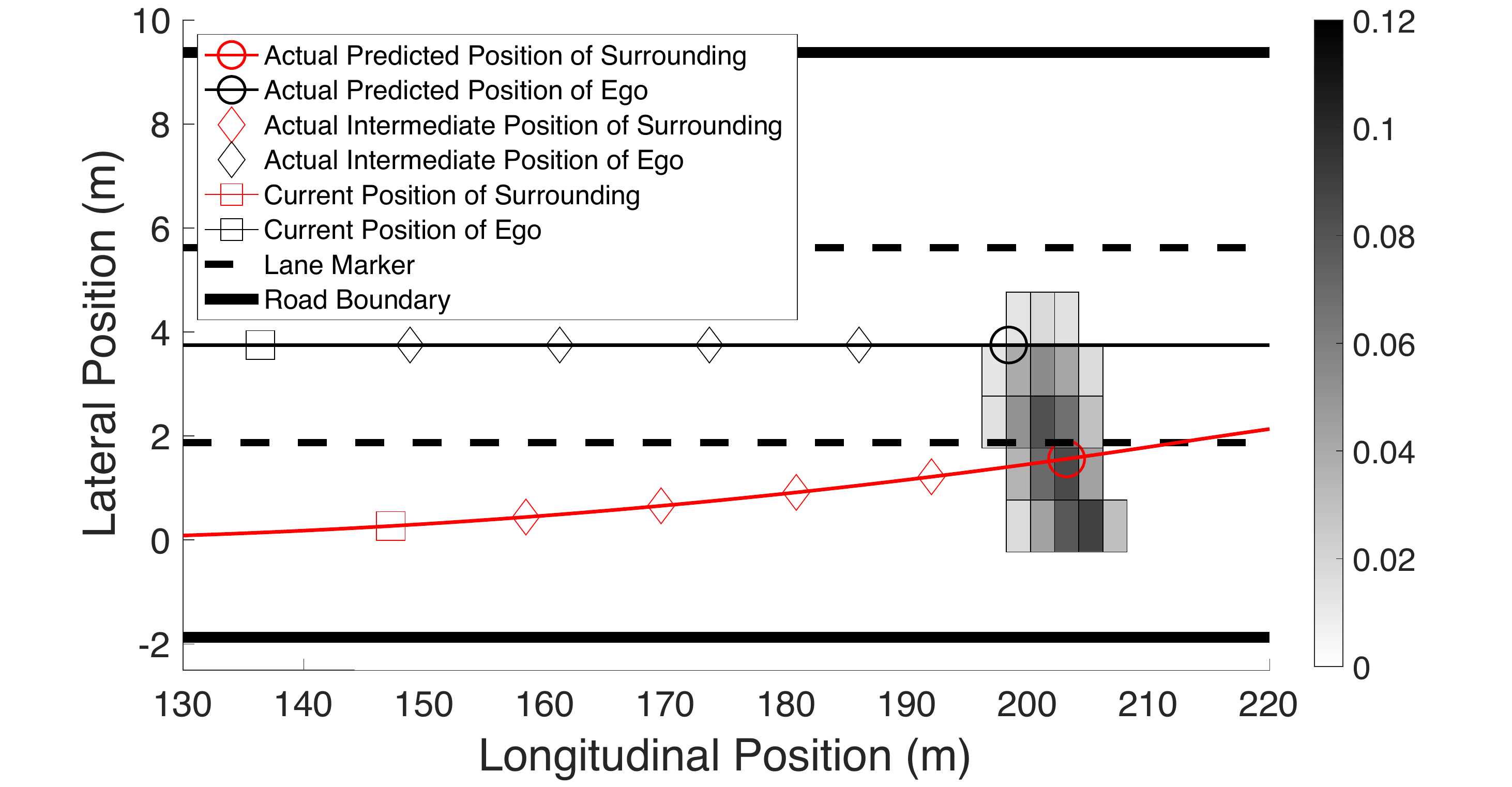}
		\caption{Visualized probability distributions  at the current time 2.4 seconds for the predicted time 4.4 seconds using the proposed approach. Only the positions with a probability greater than 1\% are displayed. Dark color indicates high distribution probability.
}\label{fig:our}
	\end{center}
\end{figure}

The visualized stochastic FRS with state probability distributions is illustrated in Fig.~\ref{fig:heu} and Fig.~\ref{fig:our} for the existing approach with default parameter settings in~\cite{althoff2009model} and the proposed prediction-based approach, respectively. At the current time $t=2.4$ seconds, the surrounding vehicle has started lane-change maneuver, and its stochastic FRS at time $t=4.4$ seconds is visualized. Note that each future predicted time step corresponds to a stochastic FRS, and we only display a single stochastic FRS at time $t=4.4$ seconds (i.e., the fifth time step) for convenience. The probabilities are aggregated for the states that share the same position with different velocities, and only the position states with greater than 1\% aggregated probabilities are plotted in the figure.

As shown in Fig.~\ref{fig:heu}, the actual future position of the surrounding vehicle is enclosed by the visualized stochastic FRS. However, the states with the highest probabilities do not overlap with the actual future position. This is because the approach in~\cite{althoff2009model} does not anticipate the lane-change maneuver of the surrounding vehicle. Although heuristic rules are employed to update the acceleration transition, the transition matrix gradually converges, leading to even probabilities for all discretized accelerations. 

Our proposed prediction-based collision detection approach indeed captures the lateral movement as shown in Fig.~\ref{fig:our}. The states around the actual future position of the surrounding vehicle have relatively higher probabilities, and more states at the left of the surrounding's current position have probabilities greater than 1\%. It indicates a more accurate collision probability estimation is realized using the proposed approach thanks to the employed prediction model.

\begin{figure}[tb]
	\begin{center}
		\includegraphics[width=0.5\textwidth]{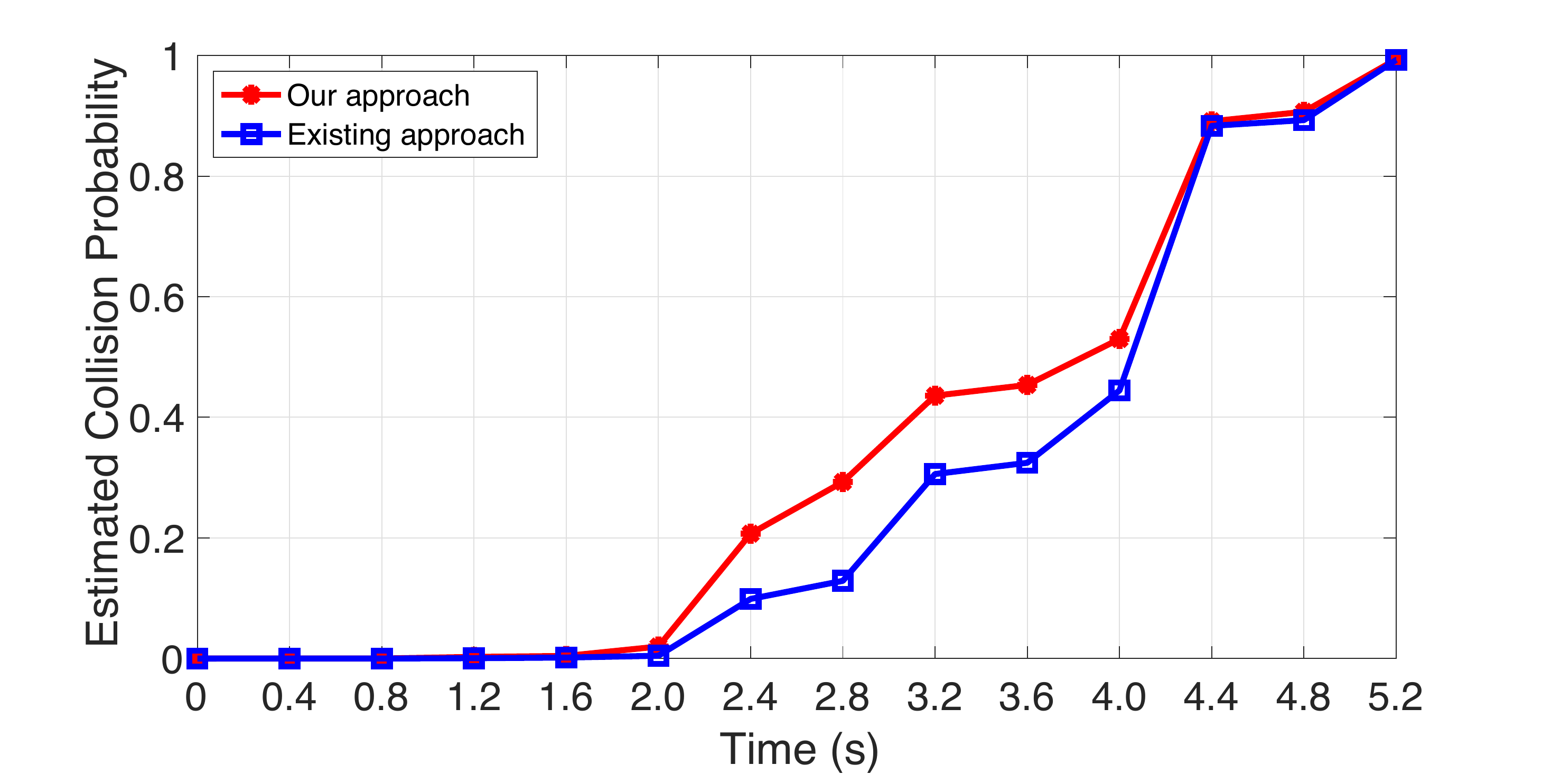}
		\caption{Estimated collision probability for a cut-in event using our approach and the existing approach in~\cite{althoff2009model}. The crash occurs at around 5 seconds. 
}\label{fig:collision}
	\end{center}
\end{figure}

We also illustrate the estimated collision probability for the simulated cut-in event in Fig.~\ref{fig:collision}. At the beginning, both two approaches measure the collision probability as zero. When the surrounding vehicle starts lane change at $t=1$ second, our approach starts to estimate the collision probability as a lower value that reaches 2.0\% at $t=2$ seconds, while the existing approach remains zero. Then the proposed approach estimates the collision probability with a sharp increase up to 20.7\% at $t=2.4$  seconds, and the collision probability further goes up afterwards. As for the existing approach, since it cannot well anticipate the cut-in maneuver, the estimated collision probability exceeds 20\% 0.8 seconds behind the proposed approach. For the last three time steps, both approaches detect a collision probability greater than 90\%, because the crash would occur soon. 

{To statistically compare the collision detection approaches,  a group of cut-in crash events is simulated as follows. We vary the ego velocity $v_e$ from 25 to 35 m/s with an 1 m/s increment, and set the surrounding vehicle velocity $v_s = v_e - v_{d}~(v_{d} = 2,3 ,4)$, resulting in entire $11 \times 3 = 33$  events. 
On average, our prediction-based approach takes 0.76 seconds less to exceed 20\% collision probability than using the baseline approach~\cite{althoff2009model}. 
{In conclusion, both approaches can identify high risks before simulated cut-in crashes, while the proposed prediction-based approach is more agile and effective.}    

\section{Conclusions}\label{sec:conc}

A highway vehicle collision detection approach leveraging prediction-based reachability analysis has been proposed in this work. The proposed approach is established on a stochastic forward reachable set, where the vehicle state probability distributions are obtained using a neural network-based acceleration prediction model. Simulation results show that the proposed prediction model can propagate 2-second vehicle positions with errors less than 0.5 meters in average. We also simulated cut-in crash events, and found that the proposed collision detection approach is more agile and effective to identify the crash thanks to the employed prediction model. Future research will {investigate more collision events, e.g., rear end crashes, and} consider infusing confidence awareness to  improve the performance of the prediction-based reachability analysis approach for collision detection and risk assessment.



%

\bibliographystyle{IEEEtran}
\bibliography{RA}

\end{document}